\documentclass[copyright]{eptcs}
\providecommand{\event}{COS~2013}
\usepackage{breakurl}

\usepackage{amsmath,latexsym,amssymb}
\usepackage[all]{xy}

\usepackage{amsthm}
\theoremstyle{plain}
\newtheorem{theorem}{Theorem}
\newtheorem{lemma}[theorem]{Lemma}

\theoremstyle{definition}
\newtheorem{definition}{Definition}

\usepackage{macrosf}

\title{Induction by Coinduction and Control Operators \\ in Call-by-Name}
\author{Yoshihiko Kakutani
 \institute{Department of Information Science, University of Tokyo, Japan}
 \email{kakutani@is.s.u-tokyo.ac.jp}
 \and
 Daisuke Kimura
 \institute{National Institute of Informatics, Japan}
 \email{kmr@nii.ac.jp}}
\def\titlerunning{Induction by Coinduction and Control Operators in Call-by-Name}
\def\authorrunning{Y.~Kakutani and D.~Kimura}

\begin{document}

\maketitle

\begin{abstract}
 This paper studies emulation of induction by coinduction
 in a call-by-name language with control operators.
 Since it is known that
 call-by-name programming languages with control operators
 cannot have general initial algebras,
 interaction of induction and control operators is often
 restricted to effect-free functions.
 We show that some class of such restricted inductive types
 can be derived from full coinductive types
 by the power of control operators.
 As a typical example of our results,
 the type of natural numbers is represented by the type of streams.
 The underlying idea is a counterpart of the fact that
 some coinductive types can be expressed by inductive types
 in call-by-name pure language without side-effects.
\end{abstract}

\section{Introduction}

In programming languages,
natural numbers are one of the most important data types.
There are many studies on natural numbers
both in the theoretic field and in the practical field.
An important feature of natural numbers is the induction principle.
In a set-theoretic approach,
the set of natural numbers $\nats$ is defined inductively.
It means that a function $H$ from $\nats$ to a set $A$
is defined by two equations
\begin{align*}
 H0 &= M \\
 H(n\nplus 1) &= F(Hn)
\end{align*}
where $M\in A$ and $F\dom{A}{A}$.
A type of natural numbers can be introduced to
the simply typed call-by-name $\lambda$-calculus,
which is core of functional programming languages, in a similar way.
Such a characterization can be regarded as an initial algebra.
For example, \cite{LambekScott86:IHOCL} is a typical study of this approach.

In a cartesian closed category with coproducts,
which is a model of the $\lambda$-calculus,
it is known that a data type of infinite streams is
the dual of the data type of natural numbers
though streams are less useful than natural numbers.
Intuitively, a stream is an infinite sequence indexed by natural numbers.
This intuition is formalized as
an isomorphism of the type of streams and
the function type from natural numbers to some data type.

The type of natural numbers can be generalized to inductive types
and the type of streams can be generalized to coinductive types.
Semantically, an inductive type is characterized as an initial algebra.
The universality of an initial algebra enables us
to define a function on the data type inductively.
Coinductive types corresponds to final coalgebras,
the dual notion of initial algebras.
In the call-by-name $\lambda$-calculus without side-effects,
a coinductive type in some class can be expressed as
a function type from an inductive type.
The construction of streams from natural numbers is
an example of this method.

The $\lambda\mu$-calculus,
an extension of the $\lambda$-calculus with first-class continuations,
was introduced by Parigot in \cite{Parigot92:LMC}.
Selinger has provided a categorical model of
the call-by-name $\lambda\mu$-calculus in \cite{Selinger01:CCD}.
It has been shown in \cite{Selinger03:SRCC} that
control operators cannot coexist with an initial object.
Since an initial object is a special case of initial algebras,
existence of general initial algebras is not allowed
in the $\lambda\mu$-calculus.
In such a situation, effect-free functions looks important.
In the second-order $\lambda\mu$-calculus,
the parametricity restricted to focal functions
was proposed by \cite{Hasegawa06:RPC}.
Focal functions are a kind of effect-free functions
and their detailed analyses are found in \cite{Selinger01:CCD}.
In this paper, we propose a reasoning about focal restriction of
initial algebras.

Unlike the case of the pure $\lambda$-calculus,
the $\lambda\mu$-calculus may not have general initial algebras.
Hence, coinductive types are not expressed by inductive types.
Instead, general coinductive types can exist in the $\lambda\mu$-calculus
and focally-restricted inductive types can be expressed
by function types from coinductive types.
This is the main result of this work.
Such contravariance between the $\lambda\mu$-calculus and
the $\lambda$-calculus derived from the CPS semantics.
Selinger's CPS transformation from the call-by-name $\lambda\mu$-calculus
to the simply typed $\lambda$-calculus
sends product types to coproduct types and
premonoidal disjunction types to product types.
This CPS transformation is based on the categorical semantics
of the $\lambda\mu$-calculus
and slightly different from Plotkin's
original CPS transformation~\cite{Plotkin75:CNCVLC}.
Barthe and Uustalu have studied a CPS transformation
with inductive types and coinductive types in \cite{BartheUustalu02:CPSTICT},
but their transformation does not deal with the universality.

The construction of this paper is the following.
In Section~\ref{SS:pre}, we prepare the target calculus.
We discuss natural numbers and streams in Section~\ref{SS:nat}.
In Section~\ref{SS:ind}, results given in Section~\ref{SS:nat} are generalized.
We show some instances of our results in Section~\ref{SS:ex}.

\section{CBN Calculus with First-Class Continuations}\label{SS:pre}

First, we introduce the base calculus, the call-by-name $\lambda\mu$-calculus.
Our version of the $\lambda\mu$-calculus is based on
Selinger's~\cite{Selinger01:CCD}.

\begin{definition}
 Types $A$, terms $M$ of the call-by-name $\lambda\mu$-calculus
 are defined by
 \begin{align*}
  A &\bnf \tau \alt A\to A \alt \top \alt A\times A
  \alt \bot \alt A\lor A \\
  M &\bnf c \alt x \alt \labst{\uptype{A}x}{M} \alt MM
  \alt \proterm \alt \pair{M}{M} \alt \proj[1]M \alt \proj[2]M
  \alt \mabst{\uptype{A}a}{M} \alt aM
  \alt \mabst{(\uptype{A}a,\uptype{A}a)}{M} \alt [a,a]M
 \end{align*}
 where $\tau$, $c$, $x$, and $a$ range over
 type constants, constants, variables and control variables, respectively.
 The constructors $\lambda$ and $\mu$ binds variables in the usual way.
 Superscripted types may be omitted.
 A judgment has the form
 \begin{align*}
  &\var{x_1}{B_1},\ldots,\var{x_n}{B_n} \prove M \type{A}
 \csep \var{a_1}{A_1},\ldots,\var{a_m}{A_m}
 \end{align*}
 with two kinds of typing environments.
 Typing rules are given in Table~\ref{FIG:lammu-type},
 where the exchange rule on environments is implicitly assumed.
 The equality is the smallest congruence relation
 including axioms in Table~\ref{FIG:lammu-eq},
 where an axiom $M \eql N$ means that
 $\Gamma \prove M \type{A} \csep \Delta$ and $\Gamma \prove N \type{A} \csep \Delta$
 are equal when both judgments are deducible.
 A substitution $\subst{a\blank}{C}$ means a recursive replacement
 of $aM$ by $C[M\subst{a\blank}{C}]$ and
 $[a_1,a_2]M$ by $C[\mabst{a}{[a_1,a_2]M\subst{a\blank}{C}}]$
 when $a$ is either $a_1$ or $a_2$.
\end{definition}

\begin{table}[t]
 \caption{Typing rules of $\lambda\mu$}
 \label{FIG:lammu-type}
 \dummy
 \begin{gather*}
  \inference{}{\Gamma \prove \uptype{A}c \type{A} \csep \Delta} \qquad
  \inference{}{\Gamma,\var{x}{A} \prove x \type{A} \csep \Delta} \qquad
  \inference{}{\Gamma \prove \proterm \type{\top} \csep \Delta} \\
  \inference{\Gamma,\var{x}{B} \prove M \type{A} \csep \Delta}
  {\Gamma \prove \labst{\uptype{B}x}{M} \type{B\to A} \csep \Delta} \qquad
  \inference{\Gamma \prove M \type{B\to A} \csep \Delta \quad
   \Gamma \prove N \type{B} \csep \Delta}
  {\Gamma \prove MN \type{A} \csep \Delta} \\
  \inference{\Gamma \prove M_1 \type{A_1} \csep \Delta \quad
   \Gamma \prove M_2 \type{A_2} \csep \Delta}
  {\Gamma \prove \pair{M_1}{M_2} \type{A_1\times A_2} \csep \Delta} \qquad
  \inference{\Gamma \prove M \type{A_1\times A_2} \csep \Delta}
  {\Gamma \prove \proj[j]M \type{A_j} \csep \Delta} \\
  \inference{\Gamma \prove M \type{\bot} \csep \var{a}{A},\Delta}
  {\Gamma \prove \mabst{\uptype{A}a}{M} \type{A} \csep \Delta} \qquad
  \inference{\Gamma \prove M \type{A} \csep \var{a}{A},\Delta}
  {\Gamma \prove aM \type{\bot} \csep \var{a}{A},\Delta} \\
  \inference{\Gamma \prove M \type{\bot} \csep \var{a_1}{A_1},\var{a_2}{A_2},\Delta}
  {\Gamma \prove \mabst{(\uptype{A_1}a_1,\uptype{A_2}a_2)}{M}
   \type{A_1\lor A_2} \csep \Delta} \qquad
  \inference{\Gamma \prove M \type{A_1\lor A_2} \csep \var{a_1}{A_1},\var{a_2}{A_2},\Delta}
  {\Gamma \prove [a_1,a_2]M \type{\bot} \csep \var{a_1}{A_1},\var{a_2}{A_2},\Delta}
 \end{gather*}
\end{table}

\begin{table}[t]
 \caption{Equality of $\lambda\mu$}
 \label{FIG:lammu-eq}
 \dummy
 \begin{align*}
  &(\labst{x}{M})N \eql M\subst{x}{N} \\
  &\labst{x}{Mx} \eql M \textif x\not\in\freev{M} \\
  &\proj[j]\pair{M_1}{M_2} \eql M_j \\
  &\pair{\proj[1]M}{\proj[2]M} \eql M \\
  &\proterm \eql M \\
  &b(\mabst{a}M) \eql M\subst{a}{b} \\
  &\mabst{a}{aM} \eql M \textif a\not\in\freev{M} \\
  &[b_1,b_2](\mabst{(a_1,a_2)}{M}) \eql M\subst{a_1,a_2}{b_1,b_2} \\
  &\mabst{(a_1,a_2)}{[a_1,a_2]M} \eql M \textif a_1,a_2\not\in\freev{M} \\
  &\uptype{\bot}bM \eql M \\
  &(\mabst{a}{M})N \eql \mabst{b}{M\subst{a\blank}{b(\blank N)}}
  \textif b\not\in\freev{M}\cup\freev{N} \\
  &\proj[j](\mabst{a}{M}) \eql \mabst{b}{M\subst{a\blank}{b(\proj[j]\blank)}}
  \textif b\not\in\freev{M} \\
  &[a_1,a_2](\mabst{a}{M}) \eql M\subst{a\blank}{[a_1,a_2]\blank}
 \end{align*}
\end{table}

Any well-typed term of the simply typed $\lambda$-calculus
without disjunction types
can be derived straightforwardly in the $\lambda\mu$-calculus
if we forget the right-hand side typing environments.
In fact, while the $\lambda$-calculus corresponds to the intuitionistic logic,
$\lambda\mu$-calculus corresponds to the classical logic.
In such Curry-Howard correspondence,
types in a left-hand side environment combine conjunctively
and types in a right-hand side environment combine disjunctively.

An equational extension of the call-by-name $\lambda\mu$-calculus
is called a $\lambda\mu$-theory.
(We also call an extension of the simply typed $\lambda$-calculus a $\lambda$-theory.)
A $\lambda\mu$-theory is needed for considering
natural numbers or other structures,
but we do not strictly distinguish a calculus from its theory
in this paper.

A $\lambda$-theory can be regarded as a category in the usual manner:
a type is an object and a function is a morphism.
In this sense, the type $\top$ is a terminal object in any $\lambda$-theory.
However, in a $\lambda\mu$-theory, the type $\bot$ is not an initial object.
In this paper, we may use the terminology of the category theory
via such translations.

For readability, $A\to\bot$ is denoted by $\lnot A$.
We also write $A_1\oplus A_2$ for $\lnot\lnot A_1\lor\lnot\lnot A_2$.
As mentioned in Selinger's note~\cite{Selinger03:SRCC},
$A_1\oplus A_2$ is more convenient for realistic programs.
We use the syntax sugar
\begin{align*}
 \inj[j]M &\equiv
 \mabst{(\uptype{\lnot\lnot A_1}a_1,\uptype{\lnot\lnot A_2}a_2)}
 a_j(\labst{\uptype{\lnot A_j}k}{kM}) \\
 \caser{F_1}{F_2} &\equiv
 \labst{\uptype{B_1\oplus B_2}x}\mabst{\uptype{A}a}{
  (\mabst{\uptype{\lnot\lnot B_2}b_2}
  {(\mabst{\uptype{\lnot\lnot B_1}b_1}{[b_1,b_2]x})(\labst{\uptype{B_1}x_1}{a(F_1x_1)})})
  (\labst{\uptype{B_2}x_2}{a(F_2x_2)})}
\end{align*}
where all introduced variables are fresh.
Then, the typing derivations
\begin{gather*}
 \inference{\Gamma \prove M \type{A_j} \csep \Delta}
 {\Gamma \prove \inj[j]M \type{A_1\oplus A_2} \csep \Delta} \qquad
 \inference{\Gamma \prove F_1 \type{B_1\to A} \csep \Delta
  \quad \Gamma \prove F_2 \type{B_2\to A} \csep \Delta}
 {\Gamma \prove \caser{F_1}{F_2} \type{B_1\oplus B_2\to A} \csep \Delta}
\end{gather*}
are admissible.
The $\beta$-like equality
\begin{align*}
 &\caser{F_1}{F_2}(\inj[j]M) \eql F_jM
\end{align*}
holds but the $\eta$-like equality
$\caser{\labst{x_1}{F(\inj[1]x_1)}}{\labst{x_2}{F(\inj[2]x_2)}} \eql F$
does not hold in general.
It follows that $\oplus$ does not give coproducts.
When $F$ is a \emph{focal} function as defined below,
the $\eta$-like equality holds.

\begin{definition}
 A term $F \type{B\to A}$ is focal
 if $F(\mabst{a}{M}) \eql \mabst{b}{M\subst{a\blank}{b(F\blank)}}$
 holds for any term $M$ and a fresh control variable $b$.
\end{definition}

A focal function can be considered an effect-free term-context
in the call-by-name $\lambda\mu$-calculus.
We define focal term-contexts $E$ as
\begin{align*}
 E &\bnf \blank \alt EM \alt \labst{x}{E} \alt \proj[1]{E} \alt \proj[2]{E}
 \alt \mabst{\uptype{A}a}{E} \alt aE
 \alt \mabst{(\uptype{A}a,\uptype{A}a)}{E} \alt [a,a]E
\end{align*}
like evaluation contexts.
Note that a focal context is not necessarily an evaluation context
because $\labst{x}{E}$ is not an evaluation context in the usual sense.
When free variables of $M$ are not captured in $E$,
we can see that
\begin{align*}
 &E[\mabst{a}{M}] \eql \mabst{b}{M\subst{a\blank}{bE}}
\end{align*}
holds.
In later sections, we often use the following fact:
for an arbitrary function $F \type{B\to A}$,
\begin{align*}
 \lnot F &\equiv \labst{\uptype{\lnot A}k}\labst{\uptype{B}x}{k(Fx)}
 &&\type{\lnot A\to\lnot B}
\end{align*}
is focal.

Categorical characterization of focal functions can be found
in \cite{Selinger01:CCD}.
In a $\lambda\mu$-theory, the notion of focal functions coincides
with the notion of central functions with respect to $\lor$:
a function $F$ is focal if and only if
$a(F(\mabst{b}{a'(G(\mabst{b'}{[b,b']x}))}))
\eql a'(G(\mabst{b'}{a(F(\mabst{b}{[b,b']x}))}))$
for any function $G$.

The notion of focal functions plays an important role
in our formulation of inductive types.
Focal functions are related to normal functions as follows.
For any term $F \type{B\to A}$,
\begin{align*}
 \focus{F} &\equiv
 \labst{\uptype{\lnot\lnot B}x}\mabst{\uptype{A}a}
 {x(\labst{\uptype{B}y}{a(Fy)})}
 &&\type{\lnot\lnot B\to A}
\end{align*}
is focal.
Moreover, this correspondence provides
bijection between functions and focal functions.
We use the notation
\begin{align*}
 \unfocus{F} &\equiv \labst{\uptype{B}x}{F(\labst{\uptype{\lnot B}k}{kx})}
\end{align*}
for the inverse transformation.

In this paper, a functor means a composition-preserving
syntactic transformation on functions of a calculus.
We assume a functor has the same codomain as its domain,
that is, a functor means an endofunctor on a calculus.
In a $\lambda\mu$-theory,
we also assume that a functor preserves focal functions.
We call a functor defined on focal functions a focus functor.
We may call also a focus functor just a functor.

\section{Natural Numbers and Streams}\label{SS:nat}

Types of natural numbers in programming languages
are studied in various ways.
A category theoretic approach is
one of the most popular characterizations of natural numbers.

In equational extensions of the simply typed $\lambda$-calculus
without side-effects,
the type of natural numbers $\nats$ with
$\zero \type{\top\to\nats}$ and $\suc \type{\nats\to\nats}$
can be defined as follows:
for any $G \type{\top\to A}$ and $F \type{A\to A}$,
there exists a unique function $H \type{\nats\to A}$ such that
each small diagram of
\begin{gather*}
 \xymatrix{
  &\top \ar[r]^-{\zero} \ar@{=}[d]
  &\nats \ar[d]^-{H}
  &\nats \ar[l]_-{\suc} \ar[d]^-{H} \\
  &\top \ar[r]^-{G}
  &A
  &A \ar[l]_-{F}
 }
\end{gather*}
commutes.
Such $H$ is denoted by $\fold[\nats]{[G,F]}$.
This natural numbers type can be considered
an initial $\functor{F}$-algebra,
where $\functor{F}$ is a functor satisfying $\functor{F}A = \top\plus A$,
if a coproduct structure $\plus$ exists in the calculus.

It is known that a model of the $\lambda\mu$-calculus becomes trivial
when it has general initial algebras.
Hence, it is not obvious that a natural numbers object
can be added to the $\lambda\mu$-calculus consistently.

On the other hand,
there exists a consistent model with general final coalgebras.
We start from adding a stream type to the $\lambda\mu$-calculus.
The type of streams $\streams$ with $\head$ and $\tail$ is defined
as the dual of natural numbers:
for any $G \type{A\to\bot}$ and $F \type{A\to A}$,
there exists a unique function $H \type{A\to\streams}$ such that
each small diagram of
\begin{gather*}
 \xymatrix{
  &\bot
  &\streams \ar[l]_-{\head} \ar[r]^-{\tail}
  &\streams \\
  &\bot \ar@{=}[u]
  &A \ar[l]_-{G} \ar[r]^-{F} \ar[u]_-{H}
  &A \ar[u]_-{H}
 }
\end{gather*}
commutes.
We write $\unfold[\streams]{\pair{G}{F}}$ for this $H$.
The type of codomain of $\head$ is not $\bot$ in usual programming,
but here $\bot$ is enough for simulating natural numbers.

We show that $\lnot\streams$ behaves like $\nats$ in a $\lambda\mu$-theory.
We define $\zero \type{\top\to\lnot\streams}$ and
$\suc \type{\lnot\streams\to\lnot\streams}$ as
\begin{align*}
 \zero &\equiv \labst{\uptype{\top}u}{\head} \\
 \suc &\equiv \labst{\uptype{\lnot\streams}y}\labst{\uptype{\streams}x}{y(\tail x)}
\end{align*}
and write $\overline{n}$ for $\suc^{n}(\zero\proterm)$.
Note that $\overline{n} \eql \head\comp\tail^n$ holds.
Let $H \type{\lnot\streams\to A}$ be
\begin{align*}
 &\labst{\uptype{\lnot\streams}y}\mabst{\uptype{A}a}
 {y((\unfold[\streams]
  {\pair{\labst{\uptype{\lnot A}h}{h(G\proterm)}}
   {\labst{\uptype{\lnot A}k}\labst{\uptype{A}x}{k(Fx)}}})
  (\labst{z}{az}))}
\end{align*}
for any $G \type{\top\to A}$ and $F \type{A\to A}$.
By the equations about $\fold[\streams]$,
equations
\begin{align*}
 H\overline{0}
 &\eql \mabst{a}{\head((\unfold[\streams]
  {\pair{\labst{h}{h(G\proterm)}}{\labst{k}\labst{x}{k(Fx)}}})
  (\labst{z}{az}))} \\
 &\eql \mabst{a}{(\labst{h}{h(G\proterm)})(\labst{z}{az})} \\
 &\eql \mabst{a}{a(G\proterm)} \eql G\proterm
\end{align*}
and
\begin{align*}
 H(\overline{n\nplus 1})
 &\eql \mabst{a}{\overline{n}(\tail((\unfold[\streams]
  {\pair{\labst{h}{h(G\proterm)}}{\labst{k}\labst{x}{k(Fx)}}})
  (\labst{z}{az})))} \\
 &\eql \mabst{a}{\overline{n}((\unfold[\streams]
  {\pair{\labst{h}{h(G\proterm)}}{\labst{k}\labst{x}{k(Fx)}}})
  ((\labst{k}\labst{x}{k(Fx)})(\labst{z}{az})))} \\
 &\eql \cdots \eql F^{n\nplus 1}(G\proterm)
\end{align*}
hold.
Since control operators are essential for the derivation,
such construction is not possible
in the pure $\lambda$-calculus without side-effects.

Unfortunately, the above $\lnot\streams$ does not satisfy
the property of $\nats$ of the $\lambda$-calculus
because $H(\suc N) \eql F(HN)$ does not necessarily hold
for arbitrary $F$ and $N$.
Instead, if $F$ is focal, $H(\suc N) \eql F(HN)$ always holds.
This fact suggests a possibility that $\lnot\streams$
may be an initial algebra for focal functions.
Since any term of the type $\top\to A$ is never focal,
we replace $\top$ with $\lnot\lnot\top$.
One can remember that functions of the type $\top\to A$
and focal functions of $\lnot\lnot\top\to A$
are in bijective correspondence.
Hence, when $F \type{A\to A}$ is focal,
\begin{gather*}
 \xymatrix{
  &\lnot\lnot\top \ar[r]^-{\focus{\zero}} \ar@{=}[d]
  &\lnot\streams \ar[d]^-{H}
  &\lnot\streams \ar[l]_-{\suc} \ar[d]^-{H} \\
  &\lnot\lnot\top \ar[r]^-{\focus{G}}
  &A
  &A \ar[l]_-{F}
 }
\end{gather*}
commutes.
Note that all functions in the diagram except for $F$
are focal without assumption and
$\focus{G}$ can range over all focal functions
from $\lnot\lnot\top$ to $A$.
Because the uniqueness of $H$ in focal functions
follows from the uniqueness of
$\unfold[\streams]{\pair{\labst{h}{h(G\proterm)}}{\labst{k}\labst{x}{k(Fx)}}}$,
the pair of $\focus{\zero}$ and $\suc$ is initial
in pairs of focal functions with the above types.

If we use the bijection $\focus{\blank}$,
we can define another type of natural numbers.
A modified type of streams $\streams'$ is defined as follows:
for any $G \type{A\to\bot}$ and $F \type{A\to\lnot\lnot A}$,
the mediating function $H = \unfold[\streams']{\pair{G}{F}}$ satisfies
\begin{gather*}
 \xymatrix{
  &\bot
  &\streams' \ar[l]_-{\head'} \ar[r]^-{\tail'}
  &\lnot\lnot\streams' \\
  &\bot \ar@{=}[u]
  &A \ar[l]_-{G} \ar[r]^-{F} \ar[u]_-{H}
  &\lnot\lnot A \ar[u]_-{\lnot\lnot H}
 }
\end{gather*}
where $\lnot\lnot H$ is
$\labst{y}\labst{k}{y(\labst{x}{k(Hx)})}$.
We can define
\begin{align*}
 \zero' &\equiv \labst{\uptype{\top}u}{\head'}
 &&\type{\top\to\lnot\streams'} \\
 \suc' &\equiv \labst{\uptype{\lnot\streams'}y}\labst{\uptype{\streams'}x}{(\tail'x)y}
 &&\type{\lnot\streams'\to\lnot\streams'}
\end{align*}
for a new type of natural numbers $\lnot\streams'$.
For a function $F \type{B\to A}$,
let $\lnot F \type{\lnot B\to\lnot A}$ be
syntax sugar of $\labst{k}\labst{x}{k(Fx)}$.
For arbitrary functions $G \type{\top\to A}$ and $F \type{A\to A}$,
there exist vertical focal functions making each small diagram of
\begin{gather*}
 \xymatrix{
  &\lnot\lnot\top \ar[r]^-{\iso} \ar@{=}[d]
  &\lnot\bot \ar[rr]^-{\lnot\head'} \ar@{=}[d] &
  &\lnot\streams' \ar[d]^-{}
  &\lnot\lnot\lnot\streams' \ar[l]_-{\lnot\tail'} \ar[d]^-{} \\
  &\lnot\lnot\top \ar[r]^-{\iso} \ar@{=}[d]
  &\lnot\bot \ar[r]^-{\iso}
  &\lnot\lnot\top \ar[r]^-{\lnot\lnot G} \ar@{=}[d]
  &\lnot\lnot A \ar[d]
  &\lnot\lnot\lnot\lnot A \ar[l]_-{\lnot\lnot\focus{F}} \ar[d] \\
  &\lnot\lnot\top \ar@{=}[rr] &
  &\lnot\lnot\top \ar[r]^-{\focus{G}}
  &A
  &\lnot\lnot A \ar[l]_-{\focus{F}}
 }
\end{gather*}
commute.
Since $\focus{(\zero')} \eql \lnot\head'\comp\mathord{\iso}$
and $\focus{(\suc')} \eql \lnot\tail'$ hold,
we can get a focal function $H$ such that the diagram
\begin{gather*}
 \xymatrix{
  &\top \ar[r]^-{\zero'} \ar@{=}[d]
  &\lnot\streams' \ar[d]^-{H}
  &\lnot\streams' \ar[l]_-{\suc'} \ar[d]^-{H} \\
  &\top \ar[r]^-{G}
  &A
  &A \ar[l]_-{F}
 }
\end{gather*}
commutes.
This $H$ is not unique in general but unique in focal functions.
While the equality for $\lnot\streams$ requires
that $F \type{A\to A}$ is focal,
the equality for $\lnot\streams'$ does not.

\section{Induction by Coinduction}\label{SS:ind}

We generalize the construction of natural numbers from streams
shown in the previous section.

First, we reformulate natural numbers as an initial algebra in focal functions.
For fixing the terminology, we give formal definitions about algebras and coalgebras.

\begin{definition}
 Let $\functor{F}$ be a functor on a $\lambda$-theory.
 A final $\functor{F}$-coalgebra is a type $C$
 with a function $X \type{C\to\functor{F}C}$ satisfying the condition:
 for any function $F \type{A\to\functor{F}A}$,
 there is a unique function $H \type{A\to C}$ such that
 \begin{gather*}
  \xymatrix{
   &\functor{F}C
   &C \ar[l]_-{X} \\
   &\functor{F}A \ar[u]_-{\functor{F}H}
   &A \ar[l]_-{F} \ar[u]_-{H}
  }
 \end{gather*}
 commutes.
 We write $\gfp[\alpha]{\functor{F}\alpha}$ for $C$
 and $\unfold{F}$ for $H$.
 An initial $\functor{F}$-algebra is the dual of a final $\functor{F}$-coalgebra.

 Let $\functor{F}$ be a focus functor on a $\lambda\mu$-theory.
 A focally initial $\functor{F}$-algebra is a type $C$
 with a focal function $X \type{\functor{F}C\to C}$ satisfying the condition:
 for any focal function $F \type{\functor{F}A\to A}$,
 there is a focal function $H \type{C\to A}$ such that
 \begin{gather*}
  \xymatrix{
   &\functor{F}C \ar[r]^-{X} \ar[d]^-{\functor{F}H}
   &C \ar[d]^-{H} \\
   &\functor{F}A \ar[r]^-{F}
   &A
  }
 \end{gather*}
 commutes and such $H$ is unique in focal functions.
 We write $\vlfp[\alpha]{\functor{F}\alpha}$ for $C$
 and $\fold{F}$ for $H$.
\end{definition}

The symbol $\mu$ is used in two meanings
for the compatibility of other studies.
We can easily classify occurrences of $\mu$:
$\mu$ in a term means a control operator
while $\mu$ in a type means a least fixed-point operator.

A focally initial $\functor{F}$-algebra means
an initial $\functor{F}$-algebra in the category of focal functions.
Our definition of focally initial algebras is
a slightly weaker notion of the definition in Hasegawa's \cite{Hasegawa06:RPC}:
in that paper, $H$ exists for non-focal $F$.

Let $\functor{F}$ be a functor such that $\functor{F}A = \lnot\lnot\top\lor A$.
The function
\begin{align*}
 &\labst{\uptype{{\lnot\lnot\top\lor\lnot\streams}}x}\mabst{\uptype{\lnot\streams}a}
 {a(\suc(\mabst{b_2}{a(\focus{\zero}(\mabst{b_1}[b_1,b_2]x))}))}
 &&\type{\functor{F}(\lnot\streams)\to\lnot\streams}
\end{align*}
is a focally initial $\functor{F}$-algebra.
We can remark that $\lor$ is not a coproduct structure in all functions
but is a coproduct structure in focal functions.
Since $\top\lor A$ is isomorphic to $\top$ in the $\lambda\mu$-calculus,
replacing $\lnot\lnot\top$ with $\top$ is meaningless
unlike the pure $\lambda$-calculus.

Let $\functor{G}$ be a functor such that $\functor{G}A = \bot\times A$.
The function
\begin{align*}
 &\labst{\uptype{\streams}x}{\pair{\head x}{\tail x}}
 &&\type{\streams\to\functor{G}\streams}
\end{align*}
is a final $\functor{G}$-coalgebra
in the usual sense.
It is the main theme of this work to investigate
when $\lnot(\gfp[\alpha]{\functor{G}\alpha})$ can be
$\vlfp[\alpha]{\functor{F}\alpha}$.

In order to generalize the construction,
we consider the CPS transformation.
CPS semantics of the call-by-name $\lambda\mu$-calculus is
provided by Selinger~\cite{Selinger01:CCD} based on
Hofmann and Streicher's transformation~\cite{HofmannStreicher97:CMULMC}.
The target language of the CPS transformation is
the simply typed $\lambda$-calculus with products and coproducts.
Let $\cps{\blank}$ be the transformation on terms
and $\cpst{\blank}$ be the transformation on types.
A judgment
\begin{align*}
 \var{x_1}{B_1},\ldots,\var{x_n}{B_n} \prove M \type{A}
 \csep \var{a_1}{A_1},\ldots,\var{a_m}{A_m}
\end{align*}
is transformed to a judgment
\begin{align*}
 \var{x_1}{\cpst{B_1}\to\ret},\ldots,\var{x_n}{\cpst{B_n}\to\ret},
 \var{a_1}{\cpst{A_1}},\ldots,\var{a_m}{\cpst{A_m}}
 \prove \cps{M} \type{\cpst{A}\to\ret}
\end{align*}
for a distinguished type $\ret$.
We do not show the details of the transformation $\cps{\blank}$ in this paper.
The type transformation is given by
\begin{align*}
 \cpst{\tau} &\equiv \tau \\
 \cpst{B\to A} &\equiv (\cpst{B}\to\ret)\times\cpst{A} \\
 \cpst{\top} &\equiv \bot \\
 \cpst{A_1\times A_2} &\equiv \cpst{A_1}\plus\cpst{A_2} \\
 \cpst{\bot} &\equiv \top \\
 \cpst{A_1\lor A_2} &\equiv \cpst{A_1}\times\cpst{A_2}
\end{align*}
when we assume the same constant types exist in the target calculus.

In the type transformation,
products in the source calculus are translated to coproducts
and premonoidal disjunctions are translated to products.
It suggests that inductive types should be translated to coinductive types
and coinductive types should be translated to inductive types
by the CPS transformation.
Indeed, $\streams = \gfp[\alpha]{\bot\times\alpha}$ can be translated to
$\lfp[\alpha]{\top\plus\alpha}$,
which is the type of natural numbers in the $\lambda$-calculus.

In a $\lambda$-theory,
it is known that $(\lfp[\alpha]{\top\plus\alpha})\to\ret$
is isomorphic to $\gfp[\alpha]{\ret\times\alpha}$.
It means intuitively that a stream is an infinite set of
elements indexed by natural numbers.
In fact, for the previous $\functor{G}$ and $\functor{F}$,
$\lnot(\gfp[\alpha]{\functor{G}\alpha}) \iso \vlfp[\alpha]{\functor{F}\alpha}$
in the source calculus is derived from
this construction in the target calculus
because $\cpst{\functor{G}A} = \top\plus\cpst{A}$ and
$\cpst{\functor{F}A} \iso \ret\times\cpst{A}$ hold.

The above construction of a coinductive type
from an inductive type in the target calculus
can be generalized as follows.
If $\functor{G}\alpha\to\ret$ is naturally isomorphic
to $\functor{F}(\alpha\to\ret)$ on $\alpha$,
then
\begin{align*}
 &(\lfp[\alpha]{\functor{G}\alpha})\to\ret
 \iso \gfp[\alpha]{\functor{F}\alpha}
\end{align*}
holds.
We lift this fact to the source calculus.

We can remark that the following property holds for the CPS transformation:
a $\lambda\mu$-term $\labst{x}{C[x]}$ is focal
if there exists a term $F$ such that $\cps{C[x]} \eql \labst{y}{x(Fy)}$ holds.
This property gives a reasoning about focal functions,
and is helpful for understanding the following theorem.

\begin{theorem}\label{THM}
 In a $\lambda\mu$-theory,
 let $\functor{F}$ be a focus functor
 and $\functor{G}$ be a functor such that
 there exists a natural isomorphism
 $\lnot\functor{G}\alpha \iso \functor{F}(\lnot\alpha)$.
 For a final $\functor{G}$-coalgebra
 $X \type{(\gfp[\alpha]{\functor{G}\alpha})\to
  \functor{G}(\gfp[\alpha]{\functor{G}\alpha})}$,
 if $X$ is focal,
 \begin{gather*}
  \xymatrix{
   &\functor{F}(\lnot(\gfp[\alpha]{\functor{G}\alpha})) \ar[r]^-{\iso}
   &\lnot\functor{G}(\gfp[\alpha]{\functor{G}\alpha}) \ar[r]^-{\lnot X}
   &\lnot(\gfp[\alpha]{\functor{G}\alpha})
  }
 \end{gather*}
 is a focally initial $\functor{F}$-algebra.
\end{theorem}

The proof of the theorem is as follows.
Given a focal function $F \type{\functor{F}A\to A}$.
Since $X$ is a final $\functor{G}$-coalgebra,
there exists $H$ such that
\begin{gather*}
 \xymatrix{
  &\functor{G}(\gfp[\alpha]{\functor{G}\alpha})
  & & & &\gfp[\alpha]{\functor{G}\alpha} \ar[llll]_-{X} \\
  &\functor{G}(\lnot A) \ar[u]_-{\functor{G}H}
  &\lnot\lnot\functor{G}(\lnot A) \ar[l]_-{M_{\functor{G}(\lnot A)}}
  &\lnot\functor{F}(\lnot\lnot A) \ar[l]_-{\iso}
  &\lnot\functor{F}A \ar[l]_-{\lnot\functor{F}M_{A}}
  &\lnot A \ar[l]_-{\lnot F} \ar[u]_-{H}
 }
\end{gather*}
commutes, where $M_{A}$ is
$\labst{\uptype{\lnot\lnot A}x}\mabst{\uptype{A}a}{x(\labst{z}{az})}$.
Let $D \type{\lnot\functor{F}A\to\functor{G}(\lnot A)}$ be
the function occurring in the bottom-line of the above diagram.
Then, the diagram
\begin{gather*}
 \xymatrix{
  &\functor{F}(\lnot(\gfp[\alpha]{\functor{G}\alpha})) \ar[r]^-{\iso}
  \ar[d]^-{\functor{F}(\lnot H)}
  &\lnot\functor{G}(\gfp[\alpha]{\functor{G}\alpha}) \ar[rr]^-{\lnot X}
  \ar[d]^-{\lnot\functor{G}H} &
  &\lnot(\gfp[\alpha]{\functor{G}\alpha}) \ar[d]^-{\lnot H} \\
  &\functor{F}(\lnot\lnot A) \ar[r]^-{\iso}
  \ar[d]^-{\functor{F}M_{A}}
  &\lnot\functor{G}(\lnot A) \ar[r]^-{\lnot D}
  &\lnot\lnot\functor{F}A \ar[r]^-{\lnot\lnot F} \ar[d]^-{M_{\functor{F}A}}
  &\lnot\lnot A \ar[d]^-{M_{A}} \\
  &\functor{F}A \ar@{=}[rr] &
  &\functor{F}A \ar[r]^-{F}
  &A
 }
\end{gather*}
commutes.
The right upper square is derived from the diagram
of the final $\functor{G}$-coalgebra.
The left upper square comes from the naturality of the isomorphism
and the right lower square means that $F$ is focal.
Though it is not trivial that the left lower square commutes,
the commuting diagram
\begin{gather*}
 \xymatrix{
  &\lnot\functor{G}(\lnot A) \ar[r]^-{\lnot M_{\functor{G}(\lnot A)}} \ar@{=}[dr]
  &\lnot\lnot\lnot\functor{G}(\lnot A) \ar[r]^-{\iso}
  \ar[d]^-{M_{\lnot\functor{G}(\lnot A)}}
  &\lnot\lnot\functor{F}(\lnot\lnot A) \ar[r]^-{\lnot\lnot\functor{F}M_{A}}
  \ar[d]^-{M_{\lnot\lnot\functor{F}A}}
  &\lnot\lnot\functor{F}A \ar[d]^-{M_{\functor{F}A}} \\
  &{}
  &\lnot\functor{G}(\lnot A) \ar[r]^-{\iso}
  &\functor{F}(\lnot\lnot A) \ar[r]^-{\functor{F}M_{A}}
  &\functor{F}A
 }
\end{gather*}
gives a proof.
Note that $\functor{F}M_{A}$ is focal because $M_{A}$ is focal
and $\functor{F}$ is a focus functor.

For a proof of the uniqueness of the mediating function,
we assume that there exists a focal function
$H' \type{\lnot(\gfp[\alpha]{\functor{G}\alpha})\to A}$
satisfying the condition.
We can have $H'' \equiv \labst{x}\mabst{a}{x(H'(\labst{z}{az}))}
\type{\lnot A\to(\gfp[\alpha]{\functor{G}\alpha})}$
such that the diagram for $X$ commutes.
The proof of its commutativity requires that $X$ is focal.
Hence, $H'' \eql H$ and $H' = M_{A}\comp\lnot H$ hold.

In the rest of the paper, we show some examples of this construction.

\section{Examples}\label{SS:ex}

We have already shown that
the type of streams $\streams$ is the most typical example.
Construction of natural numbers from another type of streams $\streams'$
is also an example of our result.
If we apply the theorem for
\begin{align*}
 \functor{F}\alpha &= \top\oplus\alpha \equiv \lnot\lnot\top\lor\lnot\lnot\alpha \\
 \functor{G}\alpha &= \lnot\top\times\lnot\lnot\alpha
\end{align*}
$\lnot\streams' \iso \lnot(\gfp[\alpha]{\lnot\top\times\lnot\lnot\alpha})$ behaves
as $\vlfp[\alpha]{\top\oplus\alpha}$.
Due to the syntax sugar,
$\lnot\streams'$ is more likely natural numbers type than $\lnot\streams$.
For arbitrary functions $G$ and $F$,
$\caser{G}{F}$ is essentially the coproduct arrow of
$\focus{G}$ and $\focus{F}$ in the category of focal functions.
Hence, in derivations of equations
\begin{align*}
 (\fold{\caser{G}{F}})(\zero'\proterm)
 &\eql (\fold{\caser{G}{F}})(\caser{\zero'}{\suc'}(\inj[1]\proterm)) \\
 &\eql \caser{G}{F}(\caser{\labst{x}{\inj[1]x}}
 {\inj[2]\comp\fold{\caser{G}{F}}}(\inj[1]\proterm)) \\
 &\eql G\proterm
\end{align*}
and
\begin{align*}
 (\fold{\caser{G}{F}})(\suc'N)
 &\eql (\fold{\caser{G}{F}})(\caser{\zero'}{\suc'}(\inj[2]N)) \\
 &\eql \caser{G}{F}(\caser{\labst{x}{\inj[1]x}}
 {\inj[2]\comp\fold{\caser{G}{F}}}(\inj[2]N)) \\
 &\eql F((\fold{\caser{G}{F}})N)
\end{align*}
focality conditions are implicit.
$\oplus$ is also useful for a list-like data type and a tree-like data type.

Before considering a list-like data type,
we consider composition of functors.
If we find two pairs of functors satisfying
the condition of Theorem~\ref{THM},
also the pair of composition functors satisfies
the condition of Theorem~\ref{THM}.

\begin{lemma}
 In a $\lambda\mu$-theory, if natural isomorphisms
 $\lnot\functor{G}_1\alpha \iso \functor{F}_1(\lnot \alpha)$ and
 $\lnot\functor{G}_2\alpha \iso \functor{F}_2(\lnot \alpha)$ exist,
 then there exists a natural isomorphism
 $\lnot\functor{G}_2\functor{G}_1\alpha \iso
 \functor{F}_2\functor{F}_1(\lnot \alpha)$.
\end{lemma}

\begin{table}[t]
 \caption{Pairs of Functors}
 \label{FIG:pairs}
 \dummy
 \begin{align*}
  &\functor{F}\alpha = \lnot\lnot\alpha &
  &\functor{G}\alpha = \lnot\lnot\alpha \\
  &\functor{F}\alpha = \lnot B\lor\alpha &
  &\functor{G}\alpha = B\times\alpha \\
  &\functor{F}\alpha = \alpha\lor\alpha &
  &\functor{G}\alpha = \alpha\times\alpha \\
  &\functor{F}\alpha = \lnot\lnot(B\times\alpha) &
  &\functor{G}\alpha = \lnot B\lor\lnot\lnot\alpha \\
  &\functor{F}\alpha = \lnot\lnot(\alpha\times\alpha) &
  &\functor{G}\alpha = \alpha\oplus\alpha
 \end{align*}
\end{table}

In Table~\ref{FIG:pairs}, we show some pairs of primitive functors
satisfying the condition of Theorem~\ref{THM}.
It is important that the connective $\times$ in the left column
is under $\lnot\lnot$.
We do not have a general idea for treating a functor
like $B\times\blank$.

In the $\lambda$-calculus without side-effects,
a type of $B$-lists is usually defined as
$\lfp[\alpha]{\top\plus(B\times\alpha)}$.
Though it is not possible to apply the theorem to
$\lfp[\alpha]{\top\lor(B\times\alpha)}$ in the $\lambda\mu$-calculus,
$\lfp[\alpha]{\top\oplus(B\times\alpha)}$ can be used.
By the lemma for compositions, the pair of functors
\begin{align*}
 \functor{F}\alpha &= \top\oplus(B\times\alpha) \\
 \functor{G}\alpha &= \lnot\top\times(\lnot B\lor\lnot\lnot\alpha)
\end{align*}
satisfies $\lnot\functor{G}\alpha \iso \functor{F}(\lnot\alpha)$.
Hence, we can get a data type of lists from a coinductive type.
If $B$ has a form $\lnot C$,
the type of $B$-lists can be expressed by
$\lnot(\gfp[\alpha]{\lnot\top\times(C\oplus\alpha)})$.

There is another difficulty caused by the fact
that $\lor$ is not monoidal but premonoidal.
Since we cannot define a full functor $\functor{G}$
so that $\functor{G}\alpha = \alpha\lor\alpha$,
$\alpha\lor\alpha$ cannot appear in the right column of the table.
On the other hand,
$\functor{F}\alpha = \alpha\lor\alpha$ is permitted in the left column
because a focus functor is required to be defined only on focal functions.
While $\alpha\lor\alpha$ is not functorial,
it can be seen that $\alpha\oplus\alpha$ is functorial.
Therefore, we can define a tree-like data type with a coinductive type.
The functors
\begin{align*}
 \functor{F}\alpha &= B\oplus(\alpha\times\alpha) \\
 \functor{G}\alpha &= \lnot B\times(\alpha\oplus\alpha)
\end{align*}
induce
$\lnot(\gfp[\alpha]{\lnot B\times(\alpha\oplus\alpha)})
\iso \vlfp[\alpha]{B\oplus(\alpha\times\alpha)}$.
It means that $\lnot(\gfp[\alpha]{\lnot B\times(\alpha\oplus\alpha)})$
behaves like a type of binary trees.
Let $\treestreams$ be $\gfp[\alpha]{\lnot B\times(\alpha\oplus\alpha)}$
and
\begin{align*}
 &\head_{\trees} \type{\treestreams\to\lnot B} \\
 &\tail_{\trees} \type{\treestreams\to\treestreams\oplus\treestreams}
\end{align*}
be functions such that $\labst{x}{\pair{\head_{\trees}x}{\tail_{\trees}x}}$
is a final $\functor{G}$-coalgebra.
For the type $\trees = \lnot\treestreams$,
\begin{align*}
 \leaf &\equiv \labst{\uptype{B}y}\labst{\uptype{\treestreams}x}
 {(\head_{\trees}x)y} \type{B\to\trees} \\
 \fork &\equiv \labst{\uptype{\trees\times\trees}y}\labst{\uptype{\treestreams}x}
 {(\mabst{a_2}{(\mabst{a_1}{[a_1,a_2](\tail_{\trees}x)})(\proj[1]y)})(\proj[2]y)}
 \type{\trees\times\trees\to\trees}
\end{align*}
can be defined.
For any functions $G \type{B\to A}$ and $F \type{A\times A\to A}$,
$\fold{\caser{G}{F}}$ is defined as
\begin{align*}
 \labst{\uptype{\trees}y}\mabst{\uptype{A}a}
 {y((\unfold(\labst{\uptype{\lnot A}k}
  {\pair{\labst{\uptype{B}x}{k(Gx)}}
   {\mabst{\uptype{\lnot A\oplus\lnot A}b}{kF'_b}}}))
  (\labst{z}{az}))}
\end{align*}
where $F'_b$ is $F\pair{\mabst{\uptype{A}b_1}{b(\inj[1](\labst{z}{b_1z}))}}
{\mabst{\uptype{A}b_2}{b(\inj[2](\labst{z}{b_2z}))}}$.
Then, we can see that two equations
\begin{align*}
 &(\fold{\caser{G}{F}})(\leaf l) \eql Gl \\
 &(\fold{\caser{G}{F}})(\fork\pair{t_1}{t_2})
 \eql F\pair{(\fold{\caser{G}{F}})t_1}{(\fold{\caser{G}{F}})t_2}
\end{align*}
hold.
Indeed, these equations mean that $\trees$ is a data type of binary trees.

\section{Concluding Remarks}

We have investigated that natural numbers can be expressed
by streams in a call-by-name language with control operators.
This is a counterpart of the fact that
a stream can be expressed by a function from natural numbers
in a pure language without side-effects.
We have generalized such method and shown construction of inductive types
from coinductive types with control operators.
As examples of the result,
a type of lists and a type of binary trees are
expressed by coinductive types.

In our study, while a coinductive type is characterized as a final coalgebra,
an inductive type is characterized as an initial algebra in focal functions.
In the $\lambda\mu$-calculus, control operators does not permit
existence of general initial algebras.
We can say that our results give
a reasoning about focal restriction of inductive types.

Our work has a possibility to connect
the duality between call-by-name and call-by-value~\cite{Filinski89:DCCD}.
In \cite{Selinger01:CCD}, it is shown that
the call-by-value $\lambda\mu$-calculus
is the dual of the call-by-name $\lambda\mu$-calculus.
Since control operators related to the classical logic,
the above duality corresponds to de~Morgan's duality.
The authors have studied the duality with
recursive structures~\cite{Kakutani02:DCNRCVI,Kimura07:CVDCNE},
and the second author has provided an extension of the dual calculus with
inductive and coinductive types~\cite{KimuraTatsuta13:CVCNDCICT}.
We are preparing another paper for the duality
extended with inductive/coinductive structures and focal restrictions.

Our construction is restricted to some class of functors.
Generalization to more complex structure with nested recursion
remains to be studied.
In the pure $\lambda$-calculus,
\cite{Altenkirch01:RFOFTTC} studies a method for more general functors
with higher-order functors.
It is not obvious to apply this method to the $\lambda\mu$-calculus
but may be helpful for generalization of our work.

\section*{Acknowledgment}

We would like to thank Kazuyuki Asada for discussing
the generalization part of this work.

\bibliographystyle{eptcs}
\bibliography{ref}

\end{document}